\documentclass[12pt]{article}

\usepackage[authoryear]{natbib}
\usepackage{amssymb,amsthm,epsfig}
\usepackage[centertags]{amsmath}
\usepackage{bm}

\newcommand{\tr}{\textrm{tr}}
\newcommand{\Es}{\mathbb{E}}

\newcommand{\diag}{\textrm{diag}}

\title{Influence diagnostics in Birnbaum--Saunders nonlinear regression models} 
\author{Artur J.~Lemonte\\   
{\small {\em Departamento de Estat\'istica, Universidade de S\~ao Paulo, Rua do Mat\~ao, 1010,}}\\
{\small {\em S\~ao Paulo/SP, 05508-090, Brazil}}}
\date{}

\begin{document}
\maketitle

\begin{abstract}

We consider the issue of assessing influence of observations in the class
of Birnbaum--Saunders nonlinear regression models, which is useful in lifetime data analysis.
Our results generalize those in Galea et al.~[2004, Influence diagnostics
in log--Birnbaum--Saunders regression models. \emph{Journal of Applied Statistics} {\bf 31}, 1049--1064]
which are confined to Birnbaum--Saunders linear regression models.
Some influence methods, such as the local influence, total local
influence of an individual and generalized leverage
are discussed. Additionally, the normal
curvatures of local influence are derived under various perturbation schemes.\\ 

\end{abstract}

\section{Introduction}

The family of distributions proposed
by Birnbaum and Saunders (1969), also known as the fatigue life distribution,
has been widely applied for describing fatigue life, and lifetimes in general.
This family of distributions was originally obtained from
a model for which failure follows from the development
and growth of a dominant crack. It was later derived by Desmond (1985) using a
biological model which followed from relaxing some of the assumptions
originally made by Birnbaum and Saunders (1969).

The random variable $T$ is said to have a Birnbaum--Saunders distribution,
say $\mathcal{B}$-$\mathcal{S}(\alpha, \eta)$,
if its density function is given by
\[
f_{T}(t; \alpha, \eta) = \frac{1}{2\alpha\eta\sqrt{2\pi}}\Biggl[\Biggl(\frac{\eta}{t}\Biggr)^{1/2}
            + \Biggl(\frac{\eta}{t}\Biggr)^{3/2}\Biggr]\exp\Biggl\{-\frac{1}{2\alpha^2}
            \Biggl(\frac{t}{\eta} + \frac{\eta}{t} - 2\Biggr)\Biggr\},\quad t>0,
\]
where $\alpha>0$ and $\eta>0$ are shape and scale parameters,
respectively. The density is right skewed, the skewness decreasing with $\alpha$.
For any $k>0$, it follows that $kT \sim\mathcal{B}$-$\mathcal{S}(\alpha, k\eta)$.
Some interesting results about improved statistical
inference for the $\mathcal{B}$-$\mathcal{S}(\alpha, \eta)$ may
be revised in Lemonte et al.~(2007, 2008). Some generalizations and extensions of the Birnbaum--Saunders
distribution are presented in \cite{Diaz-Leiva05} and \cite{GPB09}.

\cite{RiekNedelman91} proposed a log-linear regression model based on the
Birnbaum--Saunders distribution. They showed that if
$T\sim\mathcal{B}$-$\mathcal{S}(\alpha,\eta)$, then $Y=\log(T)$ is
sinh-normal distributed, say $Y\sim\mathcal{SN}(\alpha,\mu,\sigma)$,
with shape, location and scale parameters given by $\alpha$, $\mu=\log(\eta)$ and
$\sigma=2$, respectively. Diagnostic tools for the Birnbaum--Saunders
regression model were developed by \cite{Galea-etal-2004},
\cite{Leiva-etal-2007} and \cite{XiWei07}. Small-sample adjustments
for the likelihood ratio test can be found in \cite{LemFerCriCN09}.

Recently, \cite{LemCord09} proposed a new class of Birnbaum--Saunders nonlinear
regression models. The class generalizes the
regression model described by \cite{RiekNedelman91}. 
Additionally, the authors discussed maximum likelihood estimation for the parameters
of the model, and derive closed-form expressions for the
second-order biases of these estimates. 

Diagnostic analysis is an efficient way to detect
influential observations. The first technique developed to assess the
individual impact of cases on the estimation process is, perhaps, the
case deletion which became a very popular tool. 
However, case deletion excludes all information from an observation
and we can hardly say whether that observation has some influence on a
specific aspect of the model. To overcome this problem, one can resort
to local influence approach where one again investigates the model
sensibility under small perturbations. In this context,
\cite{Cook1986} proposes a general framework to detect influential
observations which give a measure of this sensibility under small
perturbations on the data or in the model. Several authors have
extended the local influence method to various regression models; see,
for example, \cite{Lawrance1988}, \cite{ThomasCook1990},
\cite{Paula1993}, \cite{LesaffreVerbeke1998}
and, more recently, \cite{Osorio-etal-2007}, 
\cite{Espinheira-etal-2008}, \cite{Paulaet-al-2009}, among others.

In this article, we present diagnostic methods
based on local influence and generalized leverage
in the class of Birnbaum--Saunders nonlinear regression models.
Our results generalize those in Galea et al.~(2004)
which are confined to Birnbaum--Saunders linear regression models.
In Section~\ref{model}, we present the class of Birnbaum--Saunders nonlinear regression models.
The score functions and observed Fisher information matrix are given as well as the process
for estimating the regression coefficients and the shape parameter.
Derivations of the normal curvature under different perturbation schemes
together with generalized leverage are made in Section~\ref{influence}.
Finally, Section~\ref{conclusion} concludes the paper.

\section{Birnbaum--Saunders nonlinear regression model}\label{model}

Let $T\sim\mathcal{B}$-$\mathcal{S}(\alpha, \eta)$. The
density function of $Y=\log(T)$ has the form
\[
\pi(y;\alpha,\mu,\sigma)=\frac{2}{\alpha\sigma\sqrt{2\pi}}\cosh\biggl(\frac{y -
\mu}{\sigma}\biggr)\exp\biggl\{-\frac{2}{\sigma^2}
\mathrm{sinh}^2\biggl(\frac{y-\mu}{\sigma}\biggr)\biggr\}, \quad y\in\mathrm{I\!R}.
\]
This distribution has a number of interesting properties:
(i) It is symmetric around the location parameter $\mu$; (ii) It is unimodal for 
$\alpha\leq 2$ and bimodal for $\alpha > 2$; (iii) $\Es(y) = \mu$ and
its variance is a function of $\alpha$ only, and has no closed-form expression,
but \cite{Rieck89} obtained asymptotic approximations for both small
and large values of $\alpha$;
(iv) If $y_{\alpha}\sim\mathcal{SN}(\alpha, \mu, \sigma)$, then
$Z_{\alpha} = 2(y_{\alpha} - \mu)/(\alpha\sigma)$ converges in distribution to the
standard normal distribution when $\alpha\to 0$.

Lemonte and Cordeiro (2009) proposed the following regression model: 
\begin{equation}\label{eq1}
y_{i}= f_{i}(\bm{x}_{i};\bm{\beta})+\varepsilon_{i},\quad i = 1,\ldots,n,
\end{equation}
where $y_{i}$ is the logarithm of the $i$th observed lifetime,
$\bm{x}_{i} = (x_{i1}, x_{i2},\ldots, x_{im})^{\top}$ is an $m\times 1$
vector of known explanatory variables
associated with the $i$th observable response $y_{i}$,
$\bm{\beta}=(\beta_1,\beta_2,\ldots,\beta_p)^{\top}$ is a vector of
unknown nonlinear parameters, and $\varepsilon_{i}\sim\mathcal{SN}(\alpha,0,2)$.
We assume a nonlinear structure for the location
parameter $\mu_{i}$ in model~(\ref{eq1}), say $\mu_{i}=f_{i}(\bm{x}_{i};\bm{\beta})$, where
$f_{i}$ is assumed to be a known and twice continuously differentiable
function with respect to $\bm{\beta}$.

The log-likelihood function for the vector parameter $\bm{\theta}=(\bm{\beta}^{\top},\alpha)^{\top}$
from a random sample $\bm{y}=(y_1,y_2,\ldots,y_n)^{\top}$ obtained from~(\ref{eq1}), can be expressed as
\begin{equation}\label{eq2}
\ell(\bm{\theta})=\sum_{i=1}^{n}\ell_{i}(\bm{\theta}),
\end{equation}
where $\ell_{i}(\bm{\theta}) = -\log(8\pi)/2 + \log(\xi_{i1}) - \xi_{i2}^{2}/2$,
\begin{equation}\label{xis}
\xi_{i1}=\xi_{i1}(\bm{\theta}) = \frac{2}{\alpha}\cosh\Bigl(\frac{y_i - \mu_{i}}{2}\Bigr),\quad
\xi_{i2}=\xi_{i2}(\bm{\theta}) = \frac{2}{\alpha}\sinh\Bigl(\frac{y_i - \mu_{i}}{2}\Bigr),
\end{equation}
for $i=1,2,\ldots,n$. The $n\times p$
local matrix $\bm{D}=\bm{D}(\bm{\beta})=\partial\bm{\mu}/\partial\bm{\beta}$
of partial derivatives of $\bm{\mu}=(\mu_{1},\mu_{2},\ldots,\mu_{n})^{\top}$ with
respect to $\bm{\beta}$ is assumed to be of full rank,
i.e., rank($\bm{D})=p$ for all $\bm{\beta}$. 

The score functions for $\bm{\beta}$ and $\alpha$ can be expressed, respectively, as
\[
\bm{U}_{\bm{\beta}} = \frac{1}{2}\bm{D}^{\top}\bm{s}\quad{\rm and}\quad
U_{\alpha}=-\frac{n}{\alpha} + \frac{1}{\alpha}\sum_{i=1}^{n}\xi_{i2}^{2},
\]
where $\bm{s}=\bm{s}(\bm{\theta})$ is an $n$-vector whose $i$th
element is equal to $\xi_{i1}\xi_{i2}-\xi_{i2}/\xi_{i1}$. 
The MLE $\widehat{\bm{\theta}} = (\widehat{\bm{\beta}}^{\top}, \widehat{\alpha})^{\top}$
satisfies $p+1$ equations: $\bm{U}_{\bm{\beta}}=\bm{0}$ and $U_{\alpha}=0$.
A joint iterative procedure to obtain the MLEs of $\bm{\beta}$ and $\alpha$
is given by (Lemonte and Cordeiro, 2009)
\[
\bm{\beta}^{(m+1)} = (\bm{D}^{(m)\top}\bm{D}^{(m)})^{-1}\bm{D}^{(m)\top}\bm{\zeta}^{(m)},\quad
\alpha^{(m+1)} = \frac{1}{2}\alpha^{(m)}(1 + \bar{\xi}_{2}^{(m)}),\quad m = 0, 1, \ldots,
\]
where $\bm{\zeta}^{(m)} = \bm{D}^{(m)}\bm{\beta}^{(m)} + \{2/\psi(\alpha^{(m)})\}\bm{s}^{(m)}$,
$\bar{\xi}_{2}^{(m)} = \sum_{i=1}^{n}\xi_{i2}^{2(m)}/n$ and
$\psi(\alpha)=2+4/\alpha^{2} - \alpha^{-1}\sqrt{2\pi}\{1 - \mathtt{erf}(\sqrt{2}/\alpha)\}
\exp(2/\alpha^2)$. Also, ${\tt erf}(\cdot)$ is the {\em error function}
(see, for example, Gradshteyn and Ryzhik, 2007).
It can be shown that $\psi(\alpha)\approx 1+4/\alpha^2$ for
$\alpha$ small and $\psi(\alpha)\approx 2$ for $\alpha$ large.
The above equations show that any software with a weighted linear regression
routine can be used to calculate the MLEs of $\bm{\beta}$ and $\alpha$
iteratively. Starting values $\bm{\beta}^{(0)}$
and $\alpha^{(0)}$ for the iterative algorithm are required.

The asymptotic inference for the parameter vector $\bm{\theta} = (\bm{\beta}^{\top},\alpha)^{\top}$
can be based on the normal approximation of the MLE of $\bm{\theta}$,
$\widehat{\bm{\theta}} = (\widehat{\bm{\beta}}^{\top},\widehat{\alpha})^{\top}$. Let
$\bm{\Sigma}_{\bm{\theta}}$ the asymptotic variance-covariance matrix
for $\widehat{\bm{\theta}}$. Then, for $n$ large,
$\widehat{\bm{\theta}}\stackrel{a}{\sim}\mathcal{N}_{p+1}(\bm{\theta},\bm{\Sigma}_{\bm{\theta}})$,
where $\stackrel{a}{\sim}$ denotes approximately distributed. Additionally,
$\bm{\Sigma}_{\bm{\theta}}$ may be approximated by
$\Ddot{\bm{L}}_{\widehat{\bm{\theta}}\widehat{\bm{\theta}}}^{-1}$, where
$\Ddot{\bm{L}}_{\widehat{\bm{\theta}}\widehat{\bm{\theta}}}$ is the $(p + 1)\times(p + 1)$
observed information matrix evaluated at $\widehat{\bm{\theta}}$, obtained from
\[
\Ddot{\bm{L}}_{\bm{\theta}\bm{\theta}} = 
\begin{bmatrix}
\Ddot{\bm{L}}_{\bm{\beta}\bm{\beta}} & \Ddot{\bm{L}}_{\bm{\beta}\alpha} \\
\Ddot{\bm{L}}_{\alpha\bm{\beta}} & \Ddot{L}_{\alpha\alpha}
\end{bmatrix}
=
\begin{bmatrix}
\bm{D}^{\top}\bm{V}\bm{D} + \frac{1}{2}[\bm{s}^{\top}][\bm{G}] & \bm{D}^{\top}\bm{h}\\
\bm{h}^{\top}\bm{D} &  \tr(\bm{K})
\end{bmatrix},
\]
where $\bm{V} = \diag\{v_1, v_2,\ldots,v_n\}$, $v_{i} = v_{i}(\bm{\theta}) =
-\{2\xi_{i2}^{2} + 4/\alpha^2 - 1 + \xi_{i2}^{2}/\xi_{i1}^{2}\}/4$,
$\bm{h} = (h_1,h_2,\ldots,h_n)^{\top}$, $h_{i} = h_{i}(\bm{\theta}) = -\xi_{i1}\xi_{i2}/\alpha$,
$\bm{K} = \diag\{k_1,k_2,\ldots,k_n\}$, $k_{i} = k_{i}(\bm{\theta}) = 1/\alpha^2 - 3\xi_{i2}^{2}/\alpha^2$ and
$\bm{G} = \bm{G}(\bm{\beta}) = \partial^2\bm{\mu}/\partial\bm{\beta}\partial\bm{\beta}^{\top}$ is
an array of dimension $n\times p\times p$. Finally, $[\cdot][\cdot]$ represents the bracket product of
a matrix by an array as defined by Wei (1998, p.~188).\footnote{If $\bm{A}$ is an
$n\times p\times q$ array and $\bm{B}$ is an $m\times n$ matrix, then $\bm{C} = [\bm{A}][\bm{B}]$
is called the bracket product of $\bm{A}$ and $\bm{B}$, that is an $m\times p\times q$ array
with elements $Y_{tij} = \sum_{k=1}^{n}B_{tk}A_{kij}$.}

\section{Diagnostic analysis}\label{influence}

\subsection{Local Influence}

The local influence method is recommended when the concern is related to investigate the model sensibility 
under some minor perturbations in the model (or data). Let $\bm{\omega}$ be a $k$-dimensional vector 
of perturbations, the perturbed log-likelihood function is denoted by $\ell(\bm{\theta}|\bm{\omega})$.
 We consider that exists a non perturbation vector, namely $\bm{\omega}_{0}$,
such that $\ell(\bm{\theta}|\bm{\omega}_{0})=\ell(\bm{\theta})$.
 The influence of minor perturbations on the maximum likelihood estimate
$\widehat{\bm{\theta}}$ can be assessed by using the log-likelihood displacement
$LD_{\bm{\omega}} = 2\{\ell(\widehat{\bm{\theta}}) - \ell(\widehat{\bm{\theta}}_{\bm{\omega}})\}$,
where $\widehat{\bm{\theta}}_{\bm{\omega}}$ denotes the maximum likelihood
estimate under $\ell(\bm{\theta}|\bm{\omega})$.

The Cook's idea for assessing local influence is essentially to analyse the local behavior of $LD_{\bm{\omega}} $
around $\bm{\omega}_{0}$ by evaluating the curvature of the plot of
$LD_{\bm{\omega}_{0} + a\bm{d}}$ against $a$, where $a\in$ I\!R and $\bm{d}$ is a unit norm direction.
 One of the measures of particular interest is the direction $\bm{d}_{\max}$ corresponding to
the largest curvature $C_{\bm{d}_{\max}}$. The index plot of $\bm{d}_{\max}$ may evidence
those observations that have considerable influence on $LD_{\bm{\omega}} $ under minor 
perturbations. Also, plots of $\bm{d}_{\max}$ against covariate values may be helpful for identifying atypical
patterns. \cite{Cook1986} shows that the normal curvature at the direction $\bm{d}$ is given by
\[
C_{\bm{d}}(\bm{\theta}) = 2|\bm{d}^{\top}\bm{\Delta}^{\top}\Ddot{\bm{L}}_{\bm{\theta}\bm{\theta}}^{-1}\bm{\Delta}\bm{d}|,
\]
where $\bm{\Delta} = \partial^2\ell(\bm{\theta}|\bm{\omega})/\partial\bm{\theta}\partial\bm{\omega}^{\top}$,
both $\bm{\Delta}$ and $\Ddot{\bm{L}}_{\bm{\theta}\bm{\theta}}$ are
evaluated at $\widehat{\bm{\theta}}$ and $\bm{\omega}_{0}$.
Hence, $C_{\bm{d}_{\max}}$ is the largest eigenvalue of
$\bm{B} = -\bm{\Delta}^{\top}\Ddot{\bm{L}}_{\bm{\theta}\bm{\theta}}^{-1}\bm{\Delta}$ and $\bm{d}_{\max}$
is the corresponding unit norm eigenvector. The index plot of $\bm{d}_{\max}$ for the matrix
$\bm{B}$ may show how to perturb the model (or data) to obtain large changes in the estimate of $\bm{\theta}$.

However, if the interest lies in computing the local influence for $\bm{\beta}$,
the normal curvature in the direction of the vector $\bm{d}$ is
$C_{\bm{d};\bm{\beta}}(\bm{\theta}) = 2|\bm{d}^{\top}\bm{\Delta}^{\top}(\Ddot{\bm{L}}_{\bm{\theta}\bm{\theta}}^{-1}-
\Ddot{\bm{L}}_{22})\bm{\Delta}\bm{d}|$, where
\[
\Ddot{\bm{L}}_{22} = 
\begin{bmatrix}
\bm{0} & \bm{0} \\
\bm{0} & \Ddot{L}_{\alpha\alpha}^{-1}
\end{bmatrix},
\]
and $\bm{d}_{\max;\bm{\beta}}$ here is the unit norm eigenvector corresponding to the largest eigenvalue of
$\bm{B}_{1} = -\bm{\Delta}^{\top}(\Ddot{\bm{L}}_{\bm{\theta}\bm{\theta}}^{-1}-\Ddot{\bm{L}}_{22})\bm{\Delta}$
\citep[see][Eq.~(26)]{Cook1986}.
The index plot of the largest eigenvector of $\bm{B}_{1}$ may reveal those influential observations on
$\widehat{\bm{\beta}}$.

Another procedure is the total local curvature corresponding to the
$i$th element, which follows by taking $\bm{d}_{i}$ or an $n\times 1$ vector of zeros with one at the $i$th
position. Thus, the curvature at the direction $\bm{d}_{i}$ assumes the form
$C_{i}(\bm{\theta}) = 2|\bm{\Delta}_{i}^{\top}\Ddot{\bm{L}}_{\bm{\theta}\bm{\theta}}^{-1}\bm{\Delta}_{i}|$,
where $\bm{\Delta}_{i}^{\top}$ denotes the $i$th
row of $\bm{\Delta}$. This is named total local influence \citep[see, for instance,][]{LesaffreVerbeke1998}.
It is also possible to compute the total local influence of the $i$th
individual when estimating a subset of the elements of $\bm{\theta}$.
For instance, if the interest lies in $\bm{\beta}$, we have that $C_{i;\bm{\beta}}(\bm{\theta})=
2|\bm{\Delta}_{i}^{\top}(\Ddot{\bm{L}}_{\bm{\theta}\bm{\theta}}^{-1}-
\Ddot{\bm{L}}_{22})\bm{\Delta}_{i}|$. Verbeke and Molembergs (2000, \S~11.3)  propose
considering as point out those cases such that $C_{i}\geq 2\bar{C}$, where $\bar{C} = \sum_{i=1}^{n}C_{i}/n$.

\subsection{Curvature calculations}

Next, we calculate, for three different perturbation scheme, the matrix
\[
\bm{\Delta} = \{\Delta_{ri}\}_{(p+1)\times n} = \biggl\{\frac{\partial^2\ell(\bm{\theta}|\bm{\omega})}
{\partial\theta_{r}\partial\omega_{i}}\biggr\},\quad r = 1,2,\ldots,p+1\quad{\rm and}\quad i = 1,2,\ldots,n,
\]
considering the model defined in~(\ref{eq1}) and its log-likelihood function given by~(\ref{eq2}).
In what follows, the quantities distinguished by the addition of ``\ $\widehat{}$\ ''
are evaluated at $\widehat{\bm{\theta}} = (\widehat{\bm{\beta}}^{\top}, \widehat{\alpha})^{\top}$.

\subsubsection{Case-weights perturbation}

The perturbation of cases is done by defining some weights for
each observation in the log-likelihood function as follows:
\[
\ell(\bm{\theta}| \bm{\omega}) = \sum_{i=1}^n \omega_i\ell_{i}(\bm{\theta}),
\]
where $\bm{\omega} = (\omega_{1},\omega_{2},\ldots,\omega_{n})^{\top}$ is the total vector of weights, 
with $0\leq\omega_{i}\leq 1$, for $i=1,2,\ldots,n$, and $\bm{\omega}_{0} = (1,1,\ldots,1)^{\top}$
is the vector of no perturbations. The matrix $\bm{\Delta}$ is given by
\[
\bm{\Delta} = 
\begin{pmatrix}
\bm{\Delta}_{\bm{\beta}} \\
\bm{\Delta}_{\alpha}
\end{pmatrix},
\]
where $\bm{\Delta}_{\bm{\beta}} =
\widehat{\bm{D}}^{\top}\diag\{\widehat{a}_{1},\widehat{a}_{2},\ldots,\widehat{a}_{n}\}$, with
$\widehat{a}_{i} = (\widehat{\xi}_{i1}\widehat{\xi}_{i2}-\widehat{\xi}_{i2}/\widehat{\xi}_{i1})/2$,
and $\bm{\Delta}_{\alpha} = (\widehat{b}_{1},\widehat{b}_{2},\ldots,\widehat{b}_{n})$,
with $\widehat{b}_{i} = -1/\widehat{\alpha} + \widehat{\xi}_{i2}^{2}/\widehat{\alpha}$. Also,
$\widehat{\xi}_{i1} = \xi_{i1}(\widehat{\bm{\theta}})$ and
$\widehat{\xi}_{i2} = \xi_{i2}(\widehat{\bm{\theta}})$, where ${\xi}_{i1}$ and ${\xi}_{i2}$
were defined in~(\ref{xis}). Note that, for linear models, the matrix $\bm{\Delta}$ reduces to the ones
given in \cite{Galea-etal-2004}.

\subsubsection{Response perturbation}\label{resp_pert}

We will consider here that each $y_i$ is perturbed as $y_{iw} = y_i + \omega_{i}S_{y}$,
where $S_{y}$ is a scale factor that may be estimated standard deviation of $\bm{y}$.
In this case, the perturbed log-likelihood function is given by
\[
\ell(\bm{\theta|\bm{\omega}}) = -\frac{n}{2}\log(8\pi) + \sum_{i=1}^{n}\log(\xi_{i1w_{1}})
-\frac{1}{2}\sum_{i=1}^{n}\xi_{i2w_{1}}^{2},
\]
where $\xi_{i1w_{1}}=\xi_{i1w_{1}}(\bm{\theta}) = 2\alpha^{-1}\cosh([y_{iw} - \mu_{i}]/2)$,
$\xi_{i2w_{1}}=\xi_{i2w_{1}}(\bm{\theta}) = 2\alpha^{-1}\sinh([y_{iw} - \mu_{i}]/2)$ and
$\bm{\omega}_{0} = (0, 0,\ldots,0)^{\top}$ is the vector of no perturbations. The matrix $\bm{\Delta}$
assumes the form
\[
\bm{\Delta} = 
\begin{pmatrix}
\bm{\Delta}_{\bm{\beta}} \\
\bm{\Delta}_{\alpha}
\end{pmatrix},
\]
where $\bm{\Delta}_{\bm{\beta}} =
\widehat{\bm{D}}^{\top}\diag\{\widehat{c}_{1},\widehat{c}_{2},\ldots,\widehat{c}_{n}\}$, with
$\widehat{c}_{i} = S_{y}(2\widehat{\xi}_{i2w_{1}}^2 + 4/\widehat{\alpha}^2 - 1
+ \widehat{\xi}_{i2w_{1}}^{2}/\widehat{\xi}_{i1w_{1}}^{2})/4$, and
$\bm{\Delta}_{\alpha} = (\widehat{d}_{1},\widehat{d}_{2},\ldots,\widehat{d}_{n})$,
with $\widehat{d}_{i} = S_{y}\widehat{\xi}_{i1w_{1}}\widehat{\xi}_{i2w_{1}}/\widehat{\alpha}$.
Also, $\widehat{\xi}_{i1w_{1}} = \xi_{i1w_{1}}(\widehat{\bm{\theta}})$ and
$\widehat{\xi}_{i2w_{1}} = \xi_{i2w_{1}}(\widehat{\bm{\theta}})$.
It is noteworthy that the matrix $\bm{\Delta}$ reduces to the ones given in \cite{Galea-etal-2004}
for linear models.

\subsubsection{Explanatory variable perturbation}

Consider now an additive perturbation on a particular continuous explanatory
variable, namely  $\bm{x}_{j}$, by making $x_{ijw} = x_{ij} + \omega_{i}S_{x}$,
where $S_{x}$ is a scale factor that may be estimated standard deviation of $\bm{x}_{j}$.
This perturbation scheme leads to the following expression for the log-likelihood function:
\[
\ell(\bm{\theta|\bm{\omega}}) = -\frac{n}{2}\log(8\pi) + \sum_{i=1}^{n}\log(\xi_{i1w_{2}})
-\frac{1}{2}\sum_{i=1}^{n}\xi_{i2w_{2}}^{2},
\]
where $\xi_{i1w_{2}}=\xi_{i1w_{2}}(\bm{\theta}) = 2\alpha^{-1}\cosh([y_{i} - \mu_{iw}]/2)$,
$\xi_{i2w_{2}}=\xi_{i2w_{2}}(\bm{\theta}) = 2\alpha^{-1}\sinh([y_{i} - \mu_{iw}]/2)$ and
$\mu_{iw} = f_{i}(\bm{x}_{iw}, \bm{\beta})$, with $\bm{x}_{iw} = (x_{i1},\ldots, x_{ijw},\ldots,x_{im})^{\top}$.
Here, $\bm{\omega}_{0} = (0, 0,\ldots,0)^{\top}$ is the vector of no perturbations.
The matrix $\bm{\Delta}$ is given by
\[
\bm{\Delta} = 
\begin{pmatrix}
\bm{\Delta}_{\bm{\beta}} \\
\bm{\Delta}_{\alpha}
\end{pmatrix},
\]
where $\bm{\Delta}_{\bm{\beta}}$ is a $p\times n$ matrix with $\Delta_{ri}$  elements that assume
the form (for $r = 1,2,\ldots,p$ and $i=1,2,\ldots,n$)
\[
\Delta_{ri} = \frac{\Ddot{\mu}_{iw}}{2}
\biggl(\widehat{\xi}_{i1w_{2}}\widehat{\xi}_{i2w_{2}}-
\frac{\widehat{\xi}_{i2w_{2}}}{\widehat{\xi}_{i1w_{2}}}\biggr)
-\frac{\dot{\mu}_{iw}\dot{\mu}_{irw}}{4}
\biggl(2\widehat{\xi}_{i2w_{2}}^2 + \frac{4}{\widehat{\alpha}^2} - 1
+ \frac{\widehat{\xi}_{i2w_{2}}^{2}}{\widehat{\xi}_{i1w_{2}}^{2}}\biggr),
\]
where $\widehat{\xi}_{i1w_{2}} = \xi_{i1w_{2}}(\widehat{\bm{\theta}})$,
$\widehat{\xi}_{i2w_{2}} = \xi_{i2w_{2}}(\widehat{\bm{\theta}})$ and
\[
\Ddot{\mu}_{iw} = \frac{\partial^2\mu_{iw}}{\partial\beta_{r}\partial\omega_{i}}
\biggr|_{\bm{\theta} = \widehat{\bm{\theta}}, \bm{\omega}=\bm{\omega}_{0}},\quad
\dot{\mu}_{iw} = \frac{\partial\mu_{iw}}{\partial\omega_{i}}
\biggr|_{\bm{\theta} = \widehat{\bm{\theta}}, \bm{\omega}=\bm{\omega}_{0}}
\quad{\rm and}\quad
\dot{\mu}_{irw} = \frac{\partial\mu_{iw}}{\partial\beta_{r}}
\biggr|_{\bm{\theta} = \widehat{\bm{\theta}}, \bm{\omega}=\bm{\omega}_{0}}.
\]
Additionally, $\bm{\Delta}_{\alpha} = (\widehat{e}_{1}, \widehat{e}_{2}, \ldots, \widehat{e}_{n})$
with $\widehat{e}_{i} = -\dot{\mu}_{iw}\widehat{\xi}_{i1w_{2}}\widehat{\xi}_{i2w_{2}}/\widehat{\alpha}$.

For linear models, i.e.~$\mu_{i} = \bm{x}_{i}^{\top}\!\bm{\beta}$,
the matrix $\bm{\Delta}_{\bm{\beta}}$ reduces to the ones given in \cite{Galea-etal-2004}.
Note that $\mu_{iw} = \bm{x}_{i}^{\top}\!\bm{\beta} + \beta_{j}w_{i}S_{x}$. Thus,
$\Ddot{\mu}_{iw} = 0$ ($r\neq j$) and $\Ddot{\mu}_{iw} = S_{x}$ ($r=j$),
$\dot{\mu}_{irw} = x_{ir}$ and $\dot{\mu}_{iw} = S_{x}\widehat{\beta}_{j}$. 
Clearly, $\bm{\Delta}_{\alpha}$ also reduces to the ones given in \cite{Galea-etal-2004} for
linear models.

\subsection{Generalized leverage}

In what follows we shall use the generalized leverage proposed by \cite{WeiHuFung1998},
which is defined as
$\bm{GL}(\widetilde{\bm{\theta}}) = \partial\widetilde{\bm{y}}/\partial\bm{y}^{\top}$,
where $\bm{\theta}$ is an $s$-vector such that $\Es(\bm{y}) = \bm{\mu}(\bm{\theta})$ and
$\widetilde{\bm{\theta}}$ is an estimator of $\bm{\theta}$, with
$\widetilde{\bm{y}} = \bm{\mu}(\widetilde{\bm{\theta}})$.
Here, the $(i, l)$ element of $\bm{GL}(\widetilde{\bm{\theta}})$, i.e.~the
generalized leverage of the estimator $\widetilde{\bm{\theta}}$ at
$(i, l)$, is the instantaneous rate of change in $i$th predicted value with respect to the $l$th
response value. As noted by the authors, the generalized leverage is invariant under
reparameterization and observations with large $GL_{ij}$ are leverage points.
\cite{WeiHuFung1998} have shown that the generalized leverage is obtained by
evaluating
\[
\bm{GL}(\bm{\theta}) = \bm{D}_{\bm{\theta}}(-\Ddot{\bm{L}}_{\bm{\theta}\bm{\theta}})^{-1}
\Ddot{\bm{L}}_{\bm{\theta}\bm{y}},
\]
at $\bm{\theta} = \widehat{\bm{\theta}}$, where
$\bm{D}_{\bm{\theta}} = \partial\bm{\mu}/\partial\bm{\theta}^{\top}$
and $\Ddot{\bm{L}}_{\bm{\theta}\bm{y}} =
\partial^2\ell(\bm{\theta})/\partial\bm{\theta}\partial\bm{y}^{\top}$.

Under model defined in~(\ref{eq1}),
we have that
\[
\bm{D}_{\bm{\theta}} = \bigl[\bm{D}\ \ \bm{0}\bigr]
\quad{\rm and}\quad
\Ddot{\bm{L}}_{\bm{\theta}\bm{y}} = -
\begin{bmatrix}
\bm{D}^{\top}\diag\{v_{1},v_{2},\ldots,v_{n}\}\\
\bm{h}
\end{bmatrix},
\]
where $v_{i}$ ($i=1,2,\ldots,n$) and $\bm{h}$ are those
as defined in Section~\ref{model}. It is noteworthy that
$\bm{GL}(\bm{\theta})$ reduces to the ones given in \cite{Galea-etal-2004} for linear models.

\section{Concluding remarks}\label{conclusion}

The Birnbaum--Saunders distribution is widely used to model times to failure for
materials subject to fatigue. In this paper, we developed
influence diagnostics for the class of Birnbaum--Saunders nonlinear regression models
which can be useful for modeling lifetime or reliability data.
Appropriate matrices for assessing local influence on the parameter
estimates under different perturbation schemes are
obtained. Our results are very general and
can be applied to any nonlinear regression model defined by~(\ref{eq1}).
In particular, our results generalize those in Galea et al.~(2004)
which are confined to Birnbaum--Saunders linear regression models.

\section*{Acknowledgments}

We gratefully acknowledge grants from FAPESP (Brazil).

\end{document}